# Efficient Privacy Preserving Logistic Regression for Horizontally Distributed Data

Guanhong Miao[a,*,1]

[a]*University of Florida, Gainesville, FL, 32611, USA*

ARTICLE INFO

*Keywords*:
Privacy preserving
Logistic regression
Matrix encryption
Chosen plaintext attack
Known plaintext attack
Collusion attack

ABSTRACT

Internet of Things devices are expanding rapidly and generating huge amount of data. There is an increasing need to explore data collected from these devices. Collaborative learning provides a strategic solution for the Internet of Things settings but also raises public concern over data privacy. In recent years, large amount of privacy preserving techniques have been developed based on secure multi-party computation and differential privacy. A major challenge of collaborative learning is to balance disclosure risk and data utility while maintaining high computation efficiency. In this paper, we proposed privacy preserving logistic regression model using matrix encryption approach. The secure scheme is resilient to chosen plaintext attack, known plaintext attack, and collusion attack that could compromise any agencies in the collaborative learning. Encrypted model estimate is decrypted to provide true model results with no accuracy degradation. Verification phase is implemented to examine dishonest behavior among agencies. Experimental evaluations demonstrate fast convergence rate and high efficiency of proposed scheme.

## 1. Introduction

Internet of Things (IoT) approaches our lives gradually with the wireless communication systems increasingly employed as technology driver for smart monitoring and applications. An IoT system can be depicted as smart devices that interact on a collaborative basis for a common goal. Smart cities are incorporating a wide range of advanced IoT infrastructures, resulting in a large amount data gathered from different IoT devices deployed in many domains, such as health care, energy transmission, transportation, and agriculture [36]. For instance, records monitored by different IoT devices can be fed to classification model for disease diagnosis in the personal health-care scenario. Smart things provide efficient tools for ubiquitous data collection or tracking, but also faces privacy threats.

In order to solve the challenges arising from IoT data process and analysis, an increasing amount of innovations have been emerged recently. With the drastically increasing amount of data generated by IoT, collaborative learning is a desirable and empowering paradigm for smart IoT systems. Collaborative learning enables multiple data providers to learn models utilizing all their data jointly [28, 19, 21]. Typical collaborative systems are distributed systems such as federated learning systems, where a central server coordinates learning process and each data provider communicates model parameters given by his own training data iteratively. Generally, models become more powerful as the training datasets grow bigger and more diverse.

Collaborative learning has benefited the society including medical research [26]. Data containing healthcare informatics is usually collected in medical centers such as hospitals. Generally, the study center does not share data with other institutes considering the confidentiality of participants. To learn disease mechanisms especially rare diseases that each study center has limited cases, it is of importance to perform data analysis combining data from multiple institutes. Collaborative learning provides great promise to connect healthcare data sources. Since data sharing of individual levels is not permitted by law or regulation in many domains, privacy preserving techniques have been developed to perform collaborative learning efficiently.

Collaborative learning with privacy preservation has been widely studied recently. Simply sharing model parameters among parties makes collaborative learning vulnerable with sensitive information being recovered [21, 9, 39, 10, 31]. Secure multi-party computation (SMC) and differentially privacy are two well-established strategies to prevent the privacy leakage. These privacy preserving methods incur either significant computational overheads or unignorable accuracy loss. Current differentially private models rarely offer acceptable privacy-utility trade-offs: settings that

(G. Miao)
ORCID(s):

provide strong privacy result in useless models, and settings with limited accuracy loss provide little effective privacy protection [14]. It is challenging to develop efficient privacy preserving technique with high accuracy.

In this paper, we propose a privacy preserving scheme for collaborative learning with high efficiency and no accuracy degradation. For a given data matrix $X$, the encrypted data is in the form of $AXB$ where $A$ is random permutation matrix and $B$ is random invertible matrix. Our contributions are as follows:

1. The proposed encryption scheme is resilient to chosen plaintext attack, known plaintext attack, and collusion attack that may compromise any agency participating collaborative learning.
2. Decryption procedure is implemented to derive true model estimate. Our scheme has no accuracy degradation with the decryption. Only model estimate is decrypted while data remain encrypted for privacy protection.
3. Verification phase is designed to check if agencies follow the encryption/decryption scheme.
4. Experiments are conducted on real data. Built on Newton's method, the proposed scheme has fast convergence rate. Moreover, the scheme is efficient to analyze large dataset.

The rest of the paper is organized as follows. Section 2 reviews the related work. Section 3 outlines preliminaries of logistic model. System overview is described in Section 4 and the construction of the privacy preserving scheme is presented in Section 5. Security analysis is given in Section 6. Section 7 provides the performance evaluation. Finally, Section 8 concludes the paper.

## 2. Related work

*Privacy preserving machine learning* Plenty privacy preserving machine learning models were investigated on distributed data. Private ridge regression using homomorphic encryption and garbled circuits was proposed in [23]. A secure linear model for collaborative learning, Helen, was proposed combining homomorphic encryption and distributed convex optimization [38]. In SecureML [22], data owners distribute their private data among two non-colluding servers who train various models on the joint data using secure two-party computation. Aggregation algorithm based on secure multi-party computation (SMC) was proposed and tailored for secure federated learning [2]. The SMC aggregation technique allows participants to encrypt their updates so that the central server only recovers the sum of the updates. There are other studies applied SMC to achieve secure distributed machine learning [33, 27]. SMC protocols tend to be highly expensive, requiring iterated encryption, decryption and repeated communication of model updates between participating parties. In SMC, unneglectable computational overhead is the main challenge for practical implementation. Moreover, unintended feature leakage from aggregated updates in collaborative learning was discussed in [21]. Recently, differentially private collaborative learning techniques were proposed to securely analyze large amount data collected by different institutes or devices [28, 24, 20, 32, 30, 15, 13]. For instance, distributed logistic regression was proposed combining differential privacy and homomorphic encryption [15]. A semi-honest server learns a differentially private model from encrypted local summary statistics at each iteration. DP-ADMM [13] was designed for distributed learning with differential privacy in which the parties exchange differentially private model parameters. The proposed algorithm is computationally efficient but requires high privacy budget to get acceptable utility of the model, with little understanding of the impact of such choices on meaningful privacy [14].

*Matrix encryption* Matrix encryption techniques have been widely studied for efficient privacy preserving data analysis. For instance, random orthogonal matrix was used to encrypt data [6] and additive noise matrix was generated to further guarantee the security. Random projection perturbation using dimension reduction approach was proposed to preserve data privacy [18]. In order to reach acceptable power, large sample size is required when analyzing encrypted data by dimension reduction approach. Elementary matrix transformation was designed to achieve secure outsourcing face recognition [35]. Moreover, secure algorithms for outsourcing linear equations and outsourcing matrix operations were investigated using matrix encryption [7, 34]. Recently, matrix filled with random integers were generated for encryption by both-sided matrix multiplication [8]. The proposed encryption approach is resilient to known plaintext attack and brute-force attack. Sparse matrix was used to design privacy preserving outsourced computation [37]. Efficient outsourcing linear regression to a cloud server was studied [5] and the vulnerability to potential disclosure attack was pointed out in [4].

## 3. Preliminaries

### 3.1. Logistic model

Consider a set of data $D = \{(\mathbf{x}_1, y_1), \cdots, (\mathbf{x}_n, y_n)\}$, where $\mathbf{x}_i \in \mathcal{R}^p$ and $y_i \in \{0, 1\}$ denotes the binary outcome such as case/control status of $\mathbf{x}_i$ $(i = 1, \cdots, n)$. The logistic regression model has the form

$$log \frac{Pr(y_i = 1|\mathbf{x}_i)}{Pr(y_i = 0|\mathbf{x}_i)} = \beta_0 + \mathbf{x}_i^T \boldsymbol{\beta}.$$

where $\boldsymbol{\beta} = (\beta_1, \cdots, \beta_p)^T$ is a $p$ dimensional coefficient vector. Model estimate of logistic regression usually fit by maximum likelihood, using the conditional likelihood. The log-likelihood for $n$ observations is

$$\ell(\beta_0, \boldsymbol{\beta}) = \sum_{i=1}^{n} \{y_i log p(\mathbf{x}_i; \beta_0, \boldsymbol{\beta}) + (1 - y_i) log(1 - p(\mathbf{x}_i; \beta_0, \boldsymbol{\beta}))\}$$

where $p(\mathbf{x}_i; \beta_0, \boldsymbol{\beta}) = Pr(y_i = 1|\mathbf{x}_i; \beta_0, \boldsymbol{\beta}) = \frac{exp(\beta_0 + \mathbf{x}_i^T \boldsymbol{\beta})}{1+exp(\beta_0 + \mathbf{x}_i^T \boldsymbol{\beta})}$.

For ridge regularized logistic regression, we maximize the log-likelihood subject to a size constraint on $L_2$-norm of the coefficients. The ridge estimate is

$$\hat{\boldsymbol{\beta}}_{ridge} = \underset{\boldsymbol{\beta}}{argmin} \{-\ell(\beta_0, \boldsymbol{\beta}) + \frac{\lambda}{2} ||\boldsymbol{\beta}||_2^2\}$$

where $\lambda \geq 0$ is the ridge parameter.

To simplify the symbols, we define $\boldsymbol{\beta} = (\beta_0, \boldsymbol{\beta})$ by including $\beta_0$ in $\boldsymbol{\beta}$. Assume the vector input $\mathbf{x}_i$ includes the constant term 1 to accommodate the intercept. Let $\mathbf{Y} = (y_1, \cdots, y_n)^T$, $\mathbf{X} = (\mathbf{x}_1, \cdots, \mathbf{x}_n)^T$ is the $n \times (p + 1)$ design matrix of the covariates, $\mathbf{p}$ is a vector of fitted probabilities with $i$-th element being $p(\mathbf{x}_i; \boldsymbol{\beta})$ and $\mathbf{W}$ is a $n \times n$ diagonal matrix of weights with $i$-th diagonal element being $p(\mathbf{x}_i; \boldsymbol{\beta})(1 - p(\mathbf{x}_i; \boldsymbol{\beta}))$.

We use the Newton's method to fit logistic regression model. Given $\boldsymbol{\beta}^{old}$, a single Newton update is

$$\boldsymbol{\beta}^{new} = \boldsymbol{\beta}^{old} - (\frac{\partial^2 \ell(\boldsymbol{\beta})}{\partial \boldsymbol{\beta} \partial \boldsymbol{\beta}^T})^{-1} \frac{\partial \ell(\boldsymbol{\beta})}{\partial \boldsymbol{\beta}}$$

where the derivatives are evaluated at $\boldsymbol{\beta}^{old}$.

$$\boldsymbol{\beta}^{new} = (\mathbf{X}^T \mathbf{W}^{old} \mathbf{X} + \boldsymbol{\Lambda})^{-1} [\mathbf{X}^T \mathbf{W}^{old} \mathbf{X} \boldsymbol{\beta}^{old} + \mathbf{X}^T (\mathbf{Y} - \mathbf{p}^{old})] \qquad (1)$$

where $\mathbf{p}^{old} = (p(\mathbf{x}_1; \boldsymbol{\beta}^{old}), \cdots, p(\mathbf{x}_n; \boldsymbol{\beta}^{old}))^T$, $p(\mathbf{x}_i; \boldsymbol{\beta}^{old})(1 - p(\mathbf{x}_i; \boldsymbol{\beta}^{old}))$ is the $i$-th diagonal element in the diagonal matrix $\mathbf{W}^{old}$, $\boldsymbol{\Lambda}$ is a matrix of 0 for regular logistic regression. $\boldsymbol{\Lambda}$ is a diagonal matrix with the diagonal elements being $\{0, \lambda, \cdots, \lambda\}$ for ridge regularized logistic regression.

## 4. System overview

### 4.1. System model

We focus on collaborative learning in which data are owned by different data providers locally. The goal is to build logistic regression model using data from all the data providers while protecting privacy.

Following the system model in [27], consider a data-driven IoT ecosystem, including IoT devices, IoT data providers, and IoT cloud server.

1. IoT devices are used for sensing and transmitting data via wireless or wired networks. The data can cover various real-world applications in smart cities, from health-related information to environmental conditions. IoT devices do not participate in the data sharing and analysis due to their limited computational capability.
2. IoT data providers collect data from IoT devices within their own domains. IoT data usually contain private information and each data provider is required to encrypt its IoT data before releasing for the analysis.
3. IoT cloud server aims at performing analysis of the encrypted data received from IoT data providers.

We define agency as data provider in this paper. Suppose there are $K$ agencies and agency $i$ holds its own data $X_i$ ($i = 1, \cdots, K$). We consider the horizontal partitioning scenario in the collaborative learning, where agencies have different sets of subjects and the same set of attributes. Agency $i$ holds its own response $Y_i$.

The proposed scheme contains pre-modeling, modeling and post-modeling phases. Data are encrypted in the pre-modeling phase. After encryption, data are sent to the cloud server to build logistic model in the modeling phase. The cloud server can be randomly selected within agencies participated in the collaborative learning. Then the cloud server sends encrypted model results back to agencies. Finally, agencies decrypt model results in the post-modeling phase.

### 4.2. Threat model

Assume the adversary model is semi-honest, i.e., agencies and the cloud server are assumed to be honest-but-curious. Precisely, agencies faithfully execute the delegated computations but may be curious about the intermediate data and try to learn or infer any sensitive information. Curious agencies are treated as inner intruders and outside intruders are assumed to have no access or know less information of the databases than participating agencies.

We design secure scheme resilient to the following adversary behaviors.

- The participating agencies may conduct chosen plaintext attack [16] to recover the encryption matrices generated by other agencies.

- If part of the original data is disclosed, the adversary can recover private data by known plaintext attack [17].

- A collusion attack may compromise any agency or cloud server.

### 4.3. Design goals

The design goals are summarized as follows.

- Privacy: We design encryption method resilient to chosen plaintext attack, known plaintext attack, and collusion attack.

- Correctness: The privacy preserving scheme is able to derive correct model result if all the agencies and cloud server behave honestly.

- Soundness: The dishonest behavior in the encryption/decryption scheme can be checked in the verification mechanism.

- Efficiency: The secure scheme is computationally efficient and achieves high accuracy.

## 5. Proposed scheme

### 5.1. Data encryption and decryption

Suppose agency $i$ has data $X_i$ with $n_i$ subjects and $p$ attributes ($i = 1, 2, \cdots, K$). The aggregated dataset from all the agencies is in the form of $X = \begin{pmatrix} X_1 \\ \vdots \\ X_K \end{pmatrix}$ and $Y = \begin{pmatrix} Y_1 \\ \vdots \\ Y_K \end{pmatrix}$. In the proposed scheme, data are encrypted in the form of $AXB$ where $B$ is an invertible matrix and $A$ is a permutation matrix. To remain data utility, $X_i$ is encrypted by all agencies with commutative encryption matrix $B_i$ (invertible encryption matrix generated by agency $i$).

**Pre-modeling phase** To make $B_i$ commutative, each agency first generates encryption key $B_0$ using the same random seed. Then agency $i$ generates a random coefficient vector $(b_{i1}, \cdots, b_{ip})$ and $B_i = \sum_{j=1}^{p} b_{ij} B_0^j$ ($1, 2, \cdots, K$). Agency $i$ generates random permutation matrix $A_{i1}, A_{i2}, \cdots, A_{iK}$ ($i = 1, 2, \cdots, K$). The dimension of $B_i$ is $p \times p$ and the dimensions of $A_{i1}, A_{i2}, \cdots, A_{iK}$ are $n_1 \times n_1, n_2 \times n_2, \cdots, n_K \times n_K$, respectively ($i = 1, 2, \cdots, K$). Agency $i$ computes $X_i^* = A_{ii} X_i B_i$ and releases $X_i^*$ to other agencies. Agency $j$ ($j \notin i$) then encrypts received data with $A_{ji}$ and $B_j$ and releases $A_{ji} X_i^* B_j$. Before sending to the cloud server, $X_i$ is encrypted by all agencies in a pre-specified order.

To protect the response information, agency $i$ computes and releases $Z_i^* = Y_i^T X_i B_i$ to other agencies. Agency $j$ ($j \notin i$) further encrypts $Z_i^*$ using the invertible matrix $B_j$. Agency $i$ also releases $B_i^T B_i$ to other agencies for

**Algorithm 1:** Pre-modeling phase

**Input:** $p \times p$ invertible matrix $B_0$ with $p$ unique eigenvalues

1 **for** *Agency* $i = 1, 2, \ldots, K$ **do**
2  generate $p$-dimensional random coefficient vector $(b_{i1}, \cdots, b_{ip})$ and permutation matrices $A_{i1}, A_{i2}, \ldots, A_{iK}$ with dimension $n_1 \times n_1, n_2 \times n_2, \ldots, n_K \times n_K$, respectively;
3  $B_i = \sum_{j=1}^{p} b_{ij} B_0^j$;

4 **for** *Agency* $i = 1, 2, \ldots, K$ **do**
5  generate $Q_i$, a permutation of $\{1, \ldots, K\}$ with $Q_i(1) = i$;
6  compute $X_i^* = A_{ii} X_i B_i$ and $Z_i^* = Y_i^T X_i B_i$, send them to $Q_i(2)$;
7  $j = 2$;
8  **while** $j \leq K$ **do**
9   Agency $Q_i(j)$ compute $X_i^* = A_{Q_i(j),i} X_i^* B_{Q_i(j)}$ and $Z_i^* = Z_i^* B_{Q_i(j)}$;
10   send $X_i^*$ and $Z_i^*$ to agency $Q_i(j+1)$ if $j < K$ or cloud server if $j = K$;
11   $j = j + 1$;

12 **if** *the goal is to build privacy preserving logistic model with ridge regularization* **then**
13  Agency 1 computes $B^* = B_1^T B_1$ and send to Agency 2;
14  **for** *Agency* $i = 2, \ldots, K$ **do**
15   compute $B^* = B_i^T B^* B_i$;
16   **if** $j \neq K$ **then**
17    send $B^*$ to Agency $i + 1$;
18   **else**
19    send $B^*$ to cloud server;

---

ridge regularized model. $B^* = \prod_{j=1}^{p} B_i^T B_i$ is the final matrix used in ridge estimate calculation. Algorithm 1 illustrates encryption procedures.

*Modeling phase* After pre-modeling phase, the cloud server receives encrypted data $X^*$, $Z_i^*$ ($i = 1, \cdots, K$) and $B^*$ for ridge regularized model. $Z_i^*$ is summed up to get a $p$ dimensional vector $Z^* = \sum_{i=1}^{K} Z_i^*$. Add a vector of 1's as the first column in $X^*$ to accommodate the intercept, i.e., $X^* = (\mathbf{1}, X^*)$. Consistently, $B_i = \begin{pmatrix} 1 & 0 \\ 0 & B_i \end{pmatrix}$ and $B^* = \begin{pmatrix} 1 & 0 \\ 0 & B^* \end{pmatrix}$. Let $\beta$ be model estimate from data $X$ and $\beta^*$ be model estimate from data $AXB$. With the encrypted data, Newton update becomes

$$\beta^{*new} = (X^{*T} W^* X^* + \Lambda B^*)^{-1} [X^{*T} W^* X^* \beta^{*old} + Z^{*T} - X^{*T} p^*] \quad (2)$$

where $\Lambda = \mathbf{0}$ for non-regularized logistic regression. The cloud server computes model estimates using equation (2) until converged.

**Theorem 1.** *The relations between $AXB$ and $X$ computation has the following four main associations. Let* $\begin{pmatrix} r(1) \\ r(2) \\ \vdots \\ r(n) \end{pmatrix}$ *be the order of $n$ samples in $X$ after permutation.*

 I. $p(x_{r(i)}^*; \beta^*) = p(x_i; \beta)$;
 II. $p^* = Ap$;
 III. $W^* = AWA^T$;

IV. $\boldsymbol{\beta}^* = \prod_{i=1}^{K} B_i^{-1} \boldsymbol{\beta}$.

*Proof.* The order of $n$ samples is permuted by the permutation matrix $A$, i.e., $\begin{pmatrix} r(1) \\ r(2) \\ \vdots \\ r(n) \end{pmatrix} = A \begin{pmatrix} 1 \\ 2 \\ \vdots \\ n \end{pmatrix}$. Let

$$P_1 = \begin{pmatrix} p(\boldsymbol{x}_1; \boldsymbol{\beta}) & & & \\ & p(\boldsymbol{x}_2; \boldsymbol{\beta}) & & \\ & & \ddots & \\ & & & p(\boldsymbol{x}_n; \boldsymbol{\beta}) \end{pmatrix}$$

and

$$P_2 = \begin{pmatrix} 1 - p(\boldsymbol{x}_1; \boldsymbol{\beta}) & & & \\ & 1 - p(\boldsymbol{x}_2; \boldsymbol{\beta}) & & \\ & & \ddots & \\ & & & 1 - p(\boldsymbol{x}_n; \boldsymbol{\beta}) \end{pmatrix}$$

where $P_1$ and $P_2$ are diagonal matrices. Let $P_1^*$ and $P_2^*$ be the corresponding matrices derived from $AXB$. So $P_1^* = A P_1 A^T$ and $P_2^* = A P_2 A^T$. $W$ can be written as

$$W = P_1 P_2.$$

Thus

$$W^* = A P_1 A^T A P_2 A^T = A W A^T.$$

For iteration $k = 0$ (initial setup), $\boldsymbol{\beta}^{(0)}$ is initial guess which is randomly generated. The selection does not affect the convergence and final model estimates.

We prove $I$-$IV$ hold in the $(m+1)$-th iteration assuming that $I$-$IV$ hold in the $m$-th iteration. To simplify notation, let $B = \prod_{i=1}^{K} B_i$ and thus $\boldsymbol{\beta}^* = B^{-1}\boldsymbol{\beta}$. Let the symbol with superscript $^{(m)}$ or subscript $_{(m)}$ denote the intermediate parameter computed in the $m$-th iteration. Given $\boldsymbol{\beta}^{(m)}$, $p(\boldsymbol{x}_i; \boldsymbol{\beta})$ is updated as $p(\boldsymbol{x}_i; \boldsymbol{\beta}^{(m)})_{(m+1)} = \frac{exp(\boldsymbol{x}_i^T \boldsymbol{\beta}^{(m)})}{1+exp(\boldsymbol{x}_i^T \boldsymbol{\beta}^{(m)})}$. Since $\boldsymbol{x}_{r(i)}^* = B^T \boldsymbol{x}_i$ and $\boldsymbol{\beta}^{*(m)} = B^{-1}\boldsymbol{\beta}^{(m)}$ (assumption $IV$ in the $m$-th iteration), we have

$$p(\boldsymbol{x}_{r(i)}^*, \boldsymbol{\beta}^{*(m)})_{(m+1)} = \frac{exp(\boldsymbol{x}_{r(i)}^{*T} \boldsymbol{\beta}^{*(m)})}{1 + exp(\boldsymbol{x}_{r(i)}^{*T} \boldsymbol{\beta}^{*(m)})}$$

$$= \frac{exp(\boldsymbol{x}_i^T B B^{-1} \boldsymbol{\beta}^{(m)})}{1 + exp(\boldsymbol{x}_i^T B B^{-1} \boldsymbol{\beta}^{(m)})}$$

$$= \frac{exp(\boldsymbol{x}_i^T \boldsymbol{\beta}^{(m)})}{1 + exp(\boldsymbol{x}_i^T \boldsymbol{\beta}^{(m)})}$$

$$= p(\boldsymbol{x}_i; \boldsymbol{\beta}^{(m)})_{(m+1)}.$$

So $I$ holds in the $(m+1)$-th iteration. It is easy to verify that $II$ and $III$ hold in the $(m+1)$-th iteration.

$$\boldsymbol{\beta}^{*(m+1)} = (X^{*T} W^{*(m)} X^* + \Lambda B^T B)^{-1}[X^{*T} W^{*(m)} X^* \boldsymbol{\beta}^{*(m)} + Z^{*T} - X^{*T} p^{*(m)}]$$

$$= (B^T X^T W^{(m)} X B + \Lambda B^T B)^{-1}[B^T X^T W^{(m)} X B B^{-1} \boldsymbol{\beta}^{(m)} + B^T Z^T - B^T X^T p^{(m)}]$$

$$= B^{-1} \boldsymbol{\beta}^{(m+1)}.$$

□

***Post-modeling phase*** After the modeling phase, the cloud server sends final results $\boldsymbol{\beta}^*$ to agencies. Agency $i$ computes $B_i\boldsymbol{\beta}^*$ and sends to other agencies for decryption ($i = 1, \cdots, K$). According to Theorem 1, $\boldsymbol{\beta} = \prod_{i=1}^{K} B_i\boldsymbol{\beta}^*$ is the final decrypted model estimate. Prediction can be performed within each agency. The probability of the outcome being 1 for $\boldsymbol{x}$ in the data is $p = 1/(1 + exp(\boldsymbol{x\beta}))$. So each agency predicts a new subject $\tilde{\boldsymbol{x}}$'s status as $1/(1 + exp(\tilde{\boldsymbol{x}}\boldsymbol{\beta}))$.

We give an example to illustrate our scheme assuming the communication order among agencies is $1 \rightarrow 2 \rightarrow \cdots \rightarrow K \rightarrow 1$ (agency $i$ sends the intermediate encrypted data to agency $i+1$, for $i < K$; agency $K$ sends data to agency 1). The final released dataset is in the form of $AXB$ where $A = \begin{pmatrix} A_{K1} \cdots A_{21}A_{11} & 0 & \cdots & 0 \\ 0 & A_{12}A_{K2} \cdots A_{22} & \cdots & 0 \\ \vdots & \vdots & \vdots & \vdots \\ 0 & 0 & \cdots & A_{(K-1)K} \cdots A_{1K}A_{KK} \end{pmatrix}$
and $B = B_1 B_2 \cdots B_K$ since $B_i$ ($i = 1, 2, \cdots, K$) is commutative. The response $\boldsymbol{Y}_i$ is contained in the encrypted dataset $\boldsymbol{Z}^* = \sum_{i=1}^{K} \boldsymbol{Y}_i^T \boldsymbol{X}_i B$. Figure 1 gives the detailed procedures of the pre-modeling and post-modeling phases.

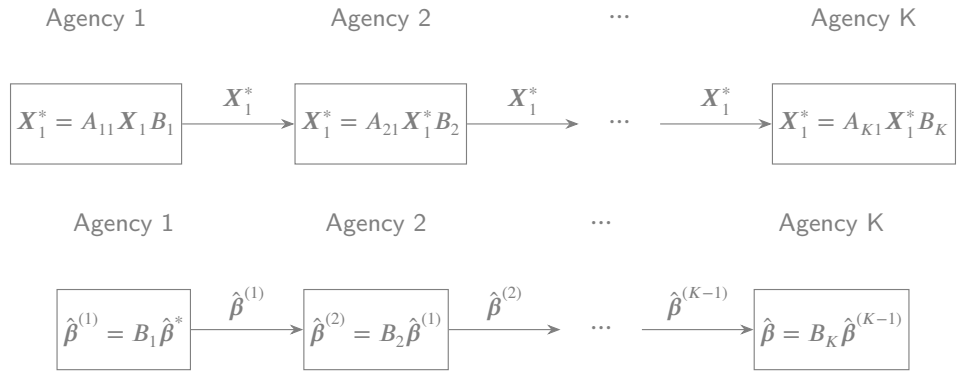

**Figure 1:** The encryption and decryption for $\boldsymbol{X}_1$ owned by agency 1. Top: encryption in the pre-modeling phase; bottom: decryption in the post-modeling phase.

### 5.2. Encryption matrix partition

For high dimensional data with large $p$, we partition the encryption key $B_0$ into diagonal matrix to reduce computation cost. For instance, there are 10,000 covariates in the dataset. If $B_0$ is partitioned with block size 100, each agency generates 100 invertible matrices with dimension $100 \times 100$ instead of one $10,000 \times 10,000$ matrix.

$$B_i = \begin{pmatrix} \tilde{B}_1 & & & \\ & \tilde{B}_2 & & \\ & & \ddots & \\ & & & \tilde{B}_{100} \end{pmatrix}$$

where $\tilde{B}_i$ ($i = 1, \cdots, 100$) is random invertible matrix generated using $B_0$. Each agency generates partitioned invertible matrix for encryption separately.

### 5.3. Cross validation

Internal cross validation is usually conducted to obtain robust model estimation. In $k$-fold cross validation, the whole dataset is partitioned into $k$ subsets. One subset is selected as the testing set and the other $k-1$ subsets are used for model training. Since data are distributed in $K$ agencies, each agency partitions its data into $k$ parts. Agency $i$ partitions $\boldsymbol{X}_i$ and $\boldsymbol{Y}_i$ into $k$ parts, defined as $\boldsymbol{X}_{it}$ and $\boldsymbol{Y}_{it}$ ($t = 1, \cdots, k$). $\boldsymbol{X}_{it}$ and $\boldsymbol{Y}_{it}^T \boldsymbol{X}_{it}$ is encrypted using Algorithm 1. The only difference is that each agency generates $k$ permutation matrices to permute the rows of $\boldsymbol{X}_{it}$ ($t = 1, \cdots, k$) respectively.

### 5.4. Verification

A verification phase is added to check if an agency performs dishonest behavior based on request. For instance, the behavior of agency $K$ is questioned and other agencies like to check if agency $K$ follows the proposed scheme honestly.

**Table 1**
Comparison of original response $Y$ and pseudo response $Y_s$.

| Response | Encrypted data | Model estimate | Purpose |
|---|---|---|---|
| $Y$ (original) | $Z_i^* = Y^T X B$ | $\beta^*$ (logistic) | Prediction |
| $Y_s$ (pseudo) | $Y_s^* = B^T X^T X \mathbf{1}$ | $\beta_s^* = (X^{*T} X^*)^{-1} Y_s^*$ | Verification |
| | $\mathbf{1} = (1, 1, \cdots, 1)^T$. | | |

To perform the verification phase, agency $i$ generates pseudo response $Y_{si} = X_i^T X_i \begin{pmatrix} 1 \\ \vdots \\ 1 \end{pmatrix}$ and encrypts it as $B_i^T Y_{si}$. Similar to the procedure used to encrypt $Y_i^T X_i$, agency $j$ ($j \neq i$) further encrypts the pseudo response as $B_j^T B_i^T Y_{si}$. The final encrypted response is in the form of $Y_{si}^* = B_1^T \cdots B_K^T Y_{si}$. Then the summation of the encrypted pseudo response (i.e., $Y_s^* = \sum_{i=1}^{K} Y_{si}^*$) is sent to the cloud server. The cloud server additionally computes $\beta_s^* = (X^{*T} X^*)^{-1} Y_s^*$ after building logistic model. So $\beta_s^* = B^{-1}(X^T X)^{-1}(B^T)^{-1} B^T Y_s = B^{-1} \begin{pmatrix} 1 \\ \vdots \\ 1 \end{pmatrix}$ where $B = B_1 \cdots B_K$ and $Y_s = \sum_{i=1}^{K} Y_{si}$.

The cloud server sends $\beta_s^*$ back to agencies in order to examine dishonest behavior. Table 1 lists details of the original and pseudo response data.

Suppose agency 1 wants to check if there is any dishonest behavior in the encryption procedure. Agency 1 computes $X_1^* \beta_s^*$ ($X_1^*$ is the final encryption matrix, i.e., $X_1^* = A_{K1} \cdots A_{11} X_1 B$). If all the agencies conduct the encryption honestly, $X_1^* \beta_s^*$ should equal $A_{*1} X_1 \begin{pmatrix} 1 \\ \vdots \\ 1 \end{pmatrix}$ where $A_{*1}$ is a permutation matrix unknown to agency 1. Agency 1 can check if there exists a permutation matrix $A_d$ such that $A_d X_1 \begin{pmatrix} 1 \\ \vdots \\ 1 \end{pmatrix} = X_1^* \beta_s^*$. If such $A_d$ exists, no dishonest behavior in the encryption procedure.

Agency 1 wants to further examine if agency $K$ follows the decryption scheme. In the post-modeling phase, agency $K$ decrypts $\beta_s^*$ using $B_K$, i.e., $\hat{\beta}^{(K-1)} = B_K \beta_s^*$. For verification, agency 1 requests the intermediate encrypted data $A_{(K-1)1} \cdots A_{11} X_1 B_1 \cdots B_{K-1}$ from other agencies. Then agency 1 computes $\beta_D = A_{*2} X_1 B_1 \cdots B_{K-1} \hat{\beta}^{(K-1)}$. If agency $K$ conducts the decryption honestly and no dishonest behaviors in encryption procedure, $\beta_D$ should equal $A_{*2} X_1 \begin{pmatrix} 1 \\ \vdots \\ 1 \end{pmatrix}$ where $A_{*2}$ is a permutation matrix unknown to agency 1. Agency 1 can check if there exists a permutation matrix $A_D$ such that $A_D X_1 \begin{pmatrix} 1 \\ \vdots \\ 1 \end{pmatrix} = \beta_D$. If no permutation matrix satisfies the equation, agency $K$ does not follow the decryption scheme. Similar approach can be applied to check if other agencies follow the decryption procedure.

## 6. Security analysis

Random permutation matrix $A$ and invertible matrix $B$ is used to perturb rows and columns of $X$. Sparse matrix encryption in the form of $AXB$ ($A$ and $B$ are sparse matrices) were investigated previously [5, 35, 7, 34, 37]. We provide security analysis for our encryption method with the framework summarized as follows.

- security of encryption matrix $B \rightarrow$ resilience of chosen plaintext attack;
- Security of $X \rightarrow$ resilience of known plaintext attack;
- Resilience of collusion attack.

## 6.1. Chosen plaintext attack

Agencies may generate fake data to perform chosen plaintext attack. We give an example to illustrate that the proposed scheme is resilient to chosen plaintext attack. Suppose agency 1 performs the attack by the following three procedures.

1. First, agency 1 sends $X_1^* = A_{11} X_1 B_1$ to agency 2.
2. Second, agency 2 encrypts $X_1^*$ as $X_{1new}^* = A_{21} X_1^* B_2$ where $A_{21}$ is random permutation matrix and $B_2 = \sum_{j=1}^{p} b_j^{(0)} B_0^j$ ($b_j^{(0)}$ is random coefficient). Then agency 2 releases the encrypted data.
3. Because only invertible encryption matrix basis $B_0$ is known by each agency, agency 1 generates random permutation matrix $A_1^+$ and $\hat{B}_2 = \sum_{j=1}^{p} b_j^* B_0^j$ ($b_j^*$ is random parameter) and then uses equation $A_{21} X_1^* B_2 = A_1^+ X_1^* \hat{B}_2$ to recover $B_2$.

Because $\hat{B}_2 = \sum_{j=1}^{p} b_j^* B_0^j$, $X_{1new}^* = A_1^+ X_1^* \hat{B}_2$ can be written as

$$X_{1new}^* = A_1^+ X_1^* (B_0 \ B_0^2 \ B_0^3 \ \cdots \ B_0^p) \begin{pmatrix} b_1^* I \\ b_2^* I \\ b_3^* I \\ \vdots \\ b_p^* I \end{pmatrix}$$

where $I$ is identity matrix. Let $U \triangleq \begin{pmatrix} b_1^* I \\ b_2^* I \\ b_3^* I \\ \vdots \\ b_p^* I \end{pmatrix}$. Each column of $U$ includes all the $p$ unknown parameters ($b_1^*, \cdots, b_p^*$) and the above equation can be broken down into $p$ sub-equations. With the $j$-th column of $U$ being the unknown vector $u_j$ and the $j$-th column of $X_{1new}^*$ being $w_j$, we have

$$w_j = A_1^+ X_1^* (B_0 \ B_0^2 \ B_0^3 \ \cdots \ B_0^p) u_j, \quad j = 1, \cdots, p.$$

Let $R \triangleq A_1^+ X_1^* (B_0 \ B_0^2 \ B_0^3 \ \cdots \ B_0^p)$ and $w_j = R u_j$. The dimension of $R$ is $n \times p^2$. To solve $u_j$, we first release the restriction of $u_j$ and do not restrict to $p$ unknown parameters. The solution is related to the rank of $R$.

1. If $rank(R) < rank([R, w_j])$, there is no solution for $u_j$.
2. If $rank(R) = rank([R, w_j]) = p^2$, there is a unique solution for $u_j$.
3. If $rank(R) = rank([R, w_j]) < p^2$, there are infinite solutions for $u_j$.

$rank(R) \leq min\{n, p\} < p^2$ because $rank(R) \leq min\{rank(A_1^+ X_1^*), rank((B_0 \ B_0^2 \ B_0^3 \ \cdots \ B_0^p))\}$. So it is impossible to have unique solution for $u_j$. If $rank(X_1^*) = min\{n, p\}$, we have $rank(R) = min\{n, p\}$ because $rank(B_0) = p$ and $rank(A_1^+) = n$. Then the solution has a direct relation with the dimension of $X_1^*$ as listed below.

1. For $n \geq p + 1$, $rank(R) = p$ and $rank([R, w_j]) = p + 1$. So there is no solution for $u_j$;
2. For $n \leq p$, $rank(R) = rank([R, w_j]) = n < p^2$. So there are infinite solutions for $u_j$.

This applies for all the $p$ sub-equations ($j = 1, \cdots, p$). Because permutation matrix for encryption is randomly generated by each agency, the true matrix $B_2$ is not a solution, i.e., $X_{1new}^* \neq A_1^+ X_1^* B_2$. A toy example is given in Appendix B. Each of the $p$ sub-equations derives different solutions of $b_j^*$ ($j = 1, \cdots, p$). So the proposed encryption scheme is resilient to chosen plaintext attack.

## 6.2. Known plaintext attack

Known plaintext attack is an effective approach [17] to recover sensitive information with both the encrypted data and partial original data released. For the proposed encryption method, we show that encryption matrices $A$ and $B$ protect against known plaintext attack. Suppose the adversary knows partial data (denoted as $X_{11}$) in the sensitive data. The first scenario (I) is $X = \begin{pmatrix} X_{11} \\ X_{22} \end{pmatrix}$ and the second scenario (II) is $X = (X_{11}, X_{22})$ where $X_{22}$ is private data.

**I.** We first give recovered data form for encrypted data with the form of $AX$ and further derive the recovered data form given encrypted data $AXB$. For $n \times p$ dimensional $X = \begin{pmatrix} X_{11} \\ X_{22} \end{pmatrix}$, suppose the encrypted data is $X^* = AX = A \begin{pmatrix} X_{11} \\ X_{22} \end{pmatrix} = \begin{pmatrix} X^*_{11} \\ X^*_{22} \end{pmatrix}$. The adversary has equation $\hat{A}^T X^* = \begin{pmatrix} X_{11} \\ \hat{X}_{22} \end{pmatrix}$ where matrices with ˆ denote recovered matrices. Because $A$ and $\hat{A}^T$ are permutation matrices, the adversary has equation

$$X^{*T}X^* = X^{*T}\hat{A}\hat{A}^T X^* = (X_{11}^T, \hat{X}_{22}^T)\begin{pmatrix} X_{11} \\ \hat{X}_{22} \end{pmatrix}.$$

The equation can be simplified as

$$X^{*T}X^* = X_{11}^T X_{11} + \hat{X}_{22}^T \hat{X}_{22}.$$

Any orthogonal transformation of $\hat{X}_{22}$ satisfying this equation can be a recovered $X_{22}$ by the adversary. Because $B$ can not be recovered (Section 6.1), we replace $X$ with $XB$ in recovered data derived above. The adversary has equation $X^{*T}X^* = X_{11}^T X_{11} + \hat{X}_{22}^T \hat{X}_{22}$ where $X^{*T}X^* = B^T[X_{11}^T X_{11} + X_{22}^T X_{22}]B$. The difference between $\hat{X}_{22}^T \hat{X}_{22}$ and $X_{22}^T X_{22}$ is $X^{*T}X^* - X_{11}^T X_{11} - X_{22}^T X_{22}$, which is also the difference between $B^T X^T X B$ and $X^T X$.

The encryption $XB = b_1 XB_0 + b_2 XB_0^2 + \cdots + b_p XB_0^p = XB_0(b_1 + b_2 B_0 + \cdots + b_p B_0^{p-1})$ can be split into two encryption steps $f_1(X) = XB_0$ and $f_2(f_1(X)) = f_1(X)(b_1 + b_2 B_0 + \cdots + b_p B_0^{p-1})$. Based on Theorem 3 (Appendix A), $B$ cannot be recovered. $b_1 + b_2 B_0 + \cdots + b_p B_0^{p-1}$ also cannot be recovered. Let $\tilde{B} = b_1 + b_2 B_0 + \cdots + b_p B_0^{p-1}$. For random invertible matrix $B_0$ with element following normal distribution $N(0, \sigma_{B_0})$, we first show that each element in $B_0^T X^T X B_0$ has an upper bound. Because $B^T X^T X B = \tilde{B}^T B_0^T X^T X B_0 \tilde{B}$, the privacy of $X^T X$ is further guaranteed by the encryption of $\tilde{B}$.

The $(i, j)$-th element in $B_0^T X^T X B_0$ (i.e., element in the $i$-th row and the $j$-th column) is the product of the $i$-th row in $B_0^T X^T$ and the $j$-th column in $XB_0$. The $j$-th column in $XB_0$ can be expressed as

$$\begin{pmatrix} x^*_{1j} \\ x^*_{2j} \\ \vdots \\ x^*_{nj} \end{pmatrix} = \begin{bmatrix} x_{11} & x_{12} & \cdots & x_{1p} \\ x_{21} & x_{22} & \cdots & x_{2p} \\ \vdots & \vdots & \vdots & \vdots \\ x_{n1} & x_{n2} & \cdots & x_{np} \end{bmatrix} \begin{pmatrix} b_{1j} \\ b_{2j} \\ \vdots \\ b_{nj} \end{pmatrix}$$

where $(b_{1j}, b_{2j}, \cdots, b_{nj})^T$ is the $j$-th column in $B_0$. Because $b_{kj}$ ($k = 1, \cdots, n$) follows normal distribution $N(0, \sigma_{B_0})$, $x^*_{cj} = \sum_{t=1}^{p} x_{ct} b_{tj}$ has normal distribution $N(0, \sigma_{B_0}^2 (\sum_{t=1}^{p} x_{ct}^2))$ for $c = 1, \cdots, n$. Because the $i$-th row in $B_0^T X^T$ is the same as the $i$-th column in $XB_0$, the $(i, j)$-th element in $B_0^T X^T X B_0$ equals $\sum_{t=1}^{n} x^*_{ti} x^*_{tj}$ where each $x^*_{ti} x^*_{tj}$ ($t \in \{1, \cdots, n\}$) follows Gamma distribution $\Gamma(1/2, 2\sigma_{B_0}^2 (\sum_{k=1}^{p} x_{tk}^2))$. Given $n$ and $X$, $\sum_{k=1}^{p} x_{tk}^2$ is fixed and $x^*_{ti} x^*_{tj} \to 0$ when $\sigma_{B_0} \to 0$. So the upper bound of the $(i, j)$-th element in $B_0^T X^T X B_0$ is adjusted by $\sigma_{B_0}$ (i.e., $\sum_{t=1}^{n} x^*_{ti} x^*_{tj} \to 0$ for $\sigma_{B_0} \to 0$). In other words, we can choose small $\sigma_{B_0}$ to increase the difference of $B_0^T X^T X B_0$ and $X^T X$ for privacy protection. The encryption of $\tilde{B}$ makes $B^T X^T X B = \tilde{B}^T B_0^T X^T X B_0 \tilde{B}$ further deviates from $X^T X$.

II. For $n \times p$ dimensional $X = (X_{11}, X_{22})$, suppose the encrypted data is $X^* = AXB = A(X_{11}, X_{22})B = (X_{11}^*, X_{22}^*)$. Assume $X^*$ and $X_{11}$ are disclosed. The columns of $X_{11}$ and $X_{22}$ are encrypted and mixed together by invertible matrix $B$. Because the adversary can not recover $B$, the adversary gets $\hat{X}_{22} = Z_{22}^*(Z_{11}^*)^{-1}X_{11}$ where $(Z_{11}^*, Z_{22}^*) = A(X_{11}, X_{22})B$. Consider a simple case where $X_{11}$ and $X_{22}$ are encrypted by separate invertible matrices $B_1$ and $B_2$ (i.e., their columns are not mixed together). The recovered $\hat{X}_{22}$ can be simplified as $\hat{X}_{22} = AX_{22}B_2[AX_{11}B_1]^{-1}X_{11} = AX_{22}B_2B_1^{-1}X_{11}^{-1}A^{-1}X_{11}$. So $X_{22}$ is still encrypted by random permutation matrix $A$ and invertible matrix $B_2$ in the recovered $\hat{X}_{22}$. This indicates that known plaintext attack is not effective for the second scenario.

Using the proposed encryption scheme, model estimate $\hat{\beta}^*$ derived by the cloud server is encrypted by invertible encryption matrix, i.e., $\hat{\beta}^* = B\hat{\beta}$ where $\hat{\beta}$ is the estimate corresponding to original data $X$. $\hat{\beta}^*$ is decrypted in the post-modeling phase (Figure 1) and prediction accuracy remains the same as non-secure model. Data $X$ remains encrypted and secure all the time.

### 6.3. Collusion attack

The proposed encryption scheme is resilient to collusion attack. Sensitive data is encrypted before releasing to other agencies. As proved above, the encryption matrix cannot be recovered. After encryption by these attack-resilient matrices, we show that the adversary cannot recover private data even when partial data is disclosed. Since any released data is accessible to all agencies, any collusion among agencies and cloud server does not provide extra information for the adversary and thus does not increase disclosure risk.

Consider a strong collusion attack that $K - 1$ out of $K$ agencies are compromised. Suppose agency 1 is the only one that remains honest. The goal of this collusion attack is to learn sensitive information of the data held by agency 1 (i.e., $X_1$). Agency 1 releases the encrypted data $A_{11}X_1B_1$ to other agencies where $A_{11}$ is a random permutation matrix and $B_1$ is invertible matrix. The adversary applies collusion attack to first recover the encryption matrices and then recover $X_1$. When agency 1 encrypts data from other agencies, the permutation encryption matrix is randomly generated for each encryption and invertible encryption matrix $B_1$ is fixed. The permutation encryption matrix can not be recovered due to its randomness. Our encryption method guarantees that $B_1$ is resilient to chosen plaintext attack and cannot be recovered by the adversary. Without prior information of $X_1$, the collusion attack fails to derive the encryption matrices and thus cannot recover $X_1$. Even partial data in $X_1$ is disclosed, our encryption method is resilient to the known plaintext attack and thus the collusion attack is not effective to derive the remaining private information. Moreover, the cloud server knows less information than agencies (i.e., data providers). Even colluding with agencies who participate the collaborative learning, the best performance of the cloud server's attack is to derive the same level of sensitive information as those compromised agencies. So both the cloud server and agencies cannot learn sensitive information in original data via the collusion attack.

## 7. Performance evaluation

We perform experiments using two datasets from the UCI repository [1]. All the experiments are performed in Matlab on University of Florida Hipergator 3.0 with 1 CPU and 4 RAMs.

*Diabetes 130-US hospitals*: This dataset was developed to identify factors related to readmission for patients with diabetes. The features were preprocessed following the procedure in [29]. A total of 69,977 samples and 42 features were used to build model predicting if the patient was readmitted within 30 days of discharge (yes/no). We use the first 60,000 samples for training and the remaining 9,977 samples for testing.

*Adult dataset*: There are 48,842 samples and 14 features in the Adult dataset. These 14 features include age, sex, education, occupation, marital status, native country, and a label representing whether the income is above $50,000 or not. Samples with missing values are excluded and the categorical features are converted to binary vectors. The label is converted to a binary outcome with $> 50k$ being 1 and $\leq 50k$ being 0. The final data contains 45,222 samples and 42 features. We use the first 40,000 samples for training and the remaining 5,222 samples for testing.

*Default of credit card clients Data Set*: There are 30,000 samples and 23 features in the Taiwan credit dataset. The goal is to predict default payment (yes/no). We use the first 25,000 samples for training and the remaining 5,000 samples for testing.

We assume the number of samples in each dataset is equally split into $K$ agencies while each subset contains all the features. The computation cost of the pre-modeling phase is increased when the number of agencies $K$ is increased.

Data utility is not related to $K$ since the total number of samples is fixed. We use the area under the receiver operating characteristic curve (AUC) to evaluate accuracy in classification problems.

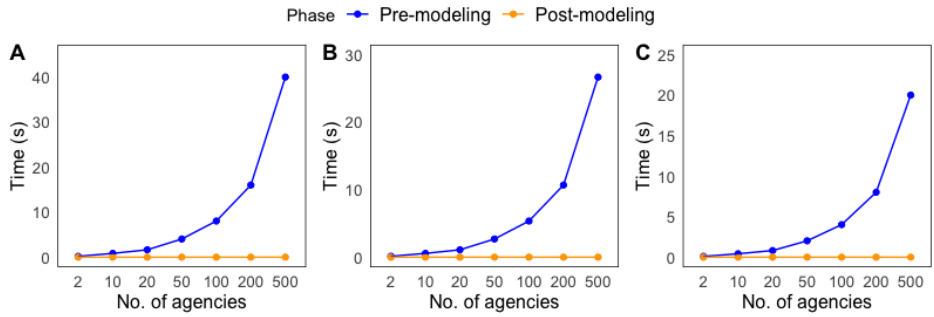

**Figure 2:** Computation cost of the encryption and decryption. A: Diabetes data; B: Adult data; C: Credit card data.

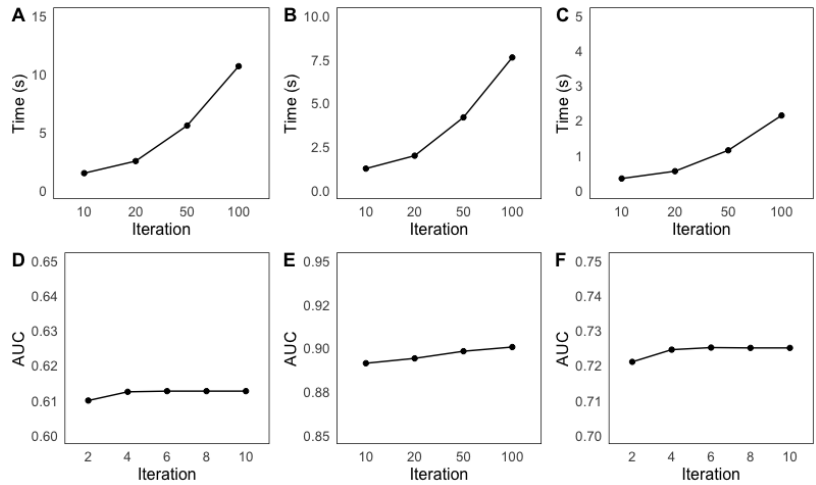

**Figure 3:** Time and AUC of the modeling phase. (A, D): Diabetes data; (B, E): Adult data; (C, F): Credit card data.

As shown in Figure 2, our privacy preserving scheme is efficient for a large number of agencies participating the collaborative learning. Based on Newton's method, our scheme converges within a small number of iterations (Figure 3). With the quadratic convergence [11], Newton's method converges faster than gradient descent [25] and alternating direction method of multipliers (ADMM) [3].

Experiments on real data show the efficiency of the proposed scheme. After 10 iterations of the Newton's method, AUCs become consistent on diabetes dataset (Figure 3D) and credit card dataset (Figure 3F). For diabetes data, the proposed secure scheme takes 2.5 seconds (1 second for pre-modeling and post-modeling phase (Figure 2) + 1.5 seconds for modeling phase (Figure 3D)) for 10 agencies to perform collaborative learning. It is more efficient than the secure logistic model proposed in [15] which takes more than 30 seconds to build logistic model for 10 participating agencies using the same dataset. Our logistic model gets AUC=0.9 within 50 iterations on adult dataset (Figure 3E). The modeling phase takes less than 5 seconds to conduct 50 iterations (Figure 3B). The computation time required by the pre-modeling and post-modeling phase is associated with the number of agencies. Experimented on the same dataset (i.e., adult dataset), our method requires less iterations to converge compared to the secure algorithm based on ADMM [13]. Although the algorithm in [13] takes less time than ours, it adds additive noise for privacy protection and is difficult to choose privacy parameter for meaningful privacy in practice [14].

## 8. Conclusion

In this paper, we propose efficient privacy preserving scheme for collaborative logistic model with data collected by different data providers. Our scheme is based on matrix encryption and resilient to chosen plaintext attack, known plaintext attack, and collusion attack that could compromise any data provider. Decryption is implemented to ensure that the proposed scheme has no accuracy degradation. The dishonest behavior deviating from the proposed scheme can be examined in the verification phase. Experiments on the real data demonstrates that our encryption scheme has fast convergence rate and is more efficient than existing secure logistic models.

## Appendix

### A. Invertible encryption matrix $B$

A polynomial $f(t)$ is said to annihilate matrix $B$ if $f(B) = 0$ (Section 3.3 of [12]). The minimal polynomial of $B$ is the monic polynomial of minimum degree that annihilates $B$. The minimal polynomial $f_B(t)$ is unique while $f_B(\lambda) = 0$ if and only if $\lambda$ is an eigenvalue of $B$. For any monic polynomial $f(t)$, $f(B) = 0$ if and only if there exists monic polynomial $h(t)$ such that $f(t) = h(t) f_B(t)$ where $f_B(t)$ is the minimal polynomial of $B$. For polynomial $f(t)$ such that $f(B) = 0$, all the eigenvalues of $B$ are the roots of $f(t)$.

**Theorem 2.** $B_i$ *is determined by* $b_{ij}$ ($j = 1, \cdots, p$) *where* $B_i$ *is generated following Algorithm 1.*

*Proof.* Let $f_1(t) = \sum_{j=1}^{p} b_{1j} t^j$, $f_2(t) = \sum_{j=1}^{p} b_{2j} t^j$ and $f(t) = f_1(t) - f_2(t)$. We have $B_1 = f_1(B_0)$ and $B_2 = f_2(B_0)$. $B_1 = B_2$ is equivalent to $f(B_0) = 0$. So $B_1 = B_2$ if and only if all the eigenvalues of $B_0$ are the roots of $f(t)$.

Suppose $\lambda_i$ ($i = 1, \cdots, p$) are the unique eigenvalues of $B_0$. In order to get $B_1 = B_2$,
$$\begin{pmatrix} \lambda_1 & \lambda_1^2 & \cdots & \lambda_1^p \\ \lambda_2 & \lambda_2^2 & \cdots & \lambda_2^p \\ \vdots & \vdots & \ddots & \vdots \\ \lambda_p & \lambda_p^2 & \cdots & \lambda_p^p \end{pmatrix} \begin{pmatrix} b_{11} \\ b_{12} \\ \vdots \\ b_{1p} \end{pmatrix} = \begin{pmatrix} \lambda_1 & \lambda_1^2 & \cdots & \lambda_1^p \\ \lambda_2 & \lambda_2^2 & \cdots & \lambda_2^p \\ \vdots & \vdots & \ddots & \vdots \\ \lambda_p & \lambda_p^2 & \cdots & \lambda_p^p \end{pmatrix} \begin{pmatrix} b_{21} \\ b_{22} \\ \vdots \\ b_{2p} \end{pmatrix}$$
symbolized by $\Lambda b^{(1)} = \Lambda b^{(2)}$. $\Lambda$ is Vandermonde matrix with rank $p$. Given $\Lambda$ and $b^{(1)}$, there is only one solution for $b^{(2)}$. So $B_1 = B_2$ if and only if $b^{(1)} = b^{(2)}$, indicating that $B_i$ is determined by $b^{(i)}$. □

**Theorem 3.** *With the knowledge of encryption key* $B_0$, *the adversary cannot recover encryption matrix* $B$.

*Proof.* Using the encryption key $B_0$ to generate commutative matrix $B$ (i.e., $B = \sum_{j=1}^{p} b_j B_0^j$), $B$ has $p$ degrees of freedom (Theorem 2). To recover $B$, the adversary needs to know these $p$ parameters (i.e., $(b_1, \cdots, b_p)$) randomly generated by the data provider (i.e., agency). So it is impossible to recover $B$. □

### B. Toy example of chosen plaintext attack

We give an example to show that the proposed encryption scheme is resilient to chosen plaintext attack. We set $n = p = 3$ and follow the procedure described in Section 6.1.

1. Generate a random invertible matrix $B_0$ with each element following normal distribution $N(0, 1)$.
2. Agency 1 generates $B_1 = 2B_0 + 3B_0^2 - 4B_0^3$, $X_1^* = IB_1$ ($I$ is identity matrix) and sends $X_1^*$ to agency 2.
3. Agency 2 encrypts $X_1^*$ as $X_{1new}^* = A_{21} X_1^* B_2$ where $A_{21}$ is random permutation matrix and $B_2 = 8B_0 + 0.3B_0^2 - 2B_0^3$.
4. Agency 1 sets identity matrix as the permutation encryption matrix (i.e., $A_1^+ = I$) and $\hat{B}_2 = b_1^* B_0 + b_2^* B_0^2 + b_3^* B_0^3$. Equation $X_{1new}^* = A_1^+ X_1^* \hat{B}_2 = X_1^* \hat{B}_2$ is used to recover $B_2$.

*Example 1* $X^*_{1new} = A_{21} X^*_1 B_2 = A_{21} I B_1 B_2$. Because $B_1 B_2 = B_2 B_1$, agency 1 gets $X^*_{1new} B_1^{-1} = A_{21} B_2 B_1 B_1^{-1} = A_{21} B_2$. Then agency 1 tries to recover $B_2$ using equation $A_{21} B_2 = \hat{B}_2 = (B_0 \; B_0^2 \; B_0^3) \begin{pmatrix} b_1^* I \\ b_2^* I \\ b_3^* I \end{pmatrix}$. It can be broken down into 3 sub-equations. Let $R = (B_0 \; B_0^2 \; B_0^3)$. The first equation is $w_1 = R u_1$ where $w_1$ is the first column of $A_{21} B_2$ and $u_1$ is the first column of $\begin{pmatrix} b_1^* I \\ b_2^* I \\ b_3^* I \end{pmatrix}$. Assume $B_0 = \begin{pmatrix} 0 & 0 & 1 \\ 0 & 1 & 0 \\ 1 & 0 & 1 \end{pmatrix}$ and $A_{21} B_2 = \begin{pmatrix} -1.7 & 0 & 4.3 \\ 0 & 6.3 & 0 \\ 4.3 & 0 & 2.6 \end{pmatrix}$. The reduced row echelon form of $(R, w_1)$ is $\begin{pmatrix} 1 & 0 & 0 & 0 & 0 & 1 & 1 & 0 & 1 & 6.0 \\ 0 & 1 & 0 & 0 & 1 & 0 & 0 & 1 & 0 & 0 \\ 0 & 0 & 1 & 1 & 0 & 1 & 1 & 0 & 2 & -1.7 \end{pmatrix}$. So the sub-equation has infinite solutions $u_1 = \begin{pmatrix} 0 \\ 0 \\ 1 \\ 1 \\ 0 \\ 0 \\ 0 \\ 0 \\ 0 \end{pmatrix} v_4 + \begin{pmatrix} 0 \\ 1 \\ 0 \\ 0 \\ 1 \\ 0 \\ 0 \\ 0 \\ 0 \end{pmatrix} v_5 + \begin{pmatrix} 1 \\ 0 \\ 1 \\ 0 \\ 0 \\ 0 \\ 1 \\ 0 \\ 0 \end{pmatrix} v_6 + \begin{pmatrix} 1 \\ 0 \\ 0 \\ 0 \\ 0 \\ 0 \\ 0 \\ 1 \\ 0 \end{pmatrix} v_7 + \begin{pmatrix} 0 \\ 1 \\ 0 \\ 0 \\ 0 \\ 0 \\ 0 \\ 0 \\ 0 \end{pmatrix} v_8 + \begin{pmatrix} 1 \\ 0 \\ 2 \\ 0 \\ 0 \\ 0 \\ 0 \\ 0 \\ 1 \end{pmatrix} v_9 + \begin{pmatrix} 6.0 \\ 0 \\ -1.7 \\ 0 \\ 0 \\ 0 \\ 0 \\ 0 \\ 0 \end{pmatrix}$ where $v_i$ ($i = 4, \cdots, 9$) can be any value. Then we add the restriction back to $u_1$ (i.e., the second, third, 5th, 6th, 8th, 9th elements equal 0). So $v_5 = v_6 = v_8 = v_9 = 0$, $v_4 = b_2^*$, $v_7 = b_3^*$, $\begin{pmatrix} 0 \\ 0 \\ 1 \end{pmatrix} v_4 + \begin{pmatrix} 1 \\ 0 \\ 1 \end{pmatrix} v_7 + \begin{pmatrix} 6.0 \\ 0 \\ -1.7 \end{pmatrix} = \begin{pmatrix} b_1^* \\ 0 \\ 0 \end{pmatrix}$. Agency 1 gets solution $(b_1^*, 6.7 - b_1^*, b_1^* - 6)$ where $b_1^*$ can be any value. Similarly, the solution for $w_2 = R u_2$ is $(b_1^*, b_2^*, b_1^* - b_2^* - 6.3)$ where $b_1^*, b_2^*$ can be any value and the solution for $w_3 = R u_3$ is $(b_1^*, 7.7 - b_1^*, b_1^* - 6)$ where $b_1^*$ can be any value. So agency 1 cannot recover $B_2$.

# References

[1] UCI machine learning repository.
[2] K. Bonawitz, V. Ivanov, B. Kreuter, A. Marcedone, H. B. McMahan, S. Patel, D. Ramage, A. Segal, and K. Seth. Practical secure aggregation for privacy-preserving machine learning. In *Proceedings of the 2017 ACM SIGSAC Conference on Computer and Communications Security*, page 1175–1191, New York, NY, USA, 2017.
[3] S. Boyd, N. Parikh, E. Chu, B. Peleato, and J. Eckstein. Distributed optimization and statistical learning via the alternating direction method of multipliers. *Found. Trends Mach. Learn.*, 3(1):1–122, 2011.
[4] Z. Cao, L. Liu, and O. Markowitch. Comment on "highly efficient linear regression outsourcing to a cloud". *IEEE Transactions on Cloud Computing*, 7(3):893–893, 2019.
[5] F. Chen, T. Xiang, X. Lei, and J. Chen. Highly efficient linear regression outsourcing to a cloud. *IEEE Transactions on Cloud Computing*, 2(4):499–508, 2014.
[6] K. Chen and L. Liu. Geometric data perturbation for privacy preserving outsourced data mining. *Knowledge and Information Systems*, 29(3):657–695, 2011.
[7] X. Chen, X. Huang, J. Li, J. Ma, W. Lou, and D. S. Wong. New algorithms for secure outsourcing of large-scale systems of linear equations. *IEEE Transactions on Information Forensics and Security*, 10(1):69–78, 2015.
[8] M. Dzwonkowski and R. Rykaczewski. Secure quaternion feistel cipher for dicom images. *IEEE Transactions on Image Processing*, 28(1):371–380, 2019.
[9] M. Fredrikson, S. Jha, and T. Ristenpart. Model inversion attacks that exploit confidence information and basic countermeasures. In *Proceedings of the 22nd ACM SIGSAC Conference on Computer and Communications Security*, page 1322–1333, New York, NY, USA, 2015. Association for Computing Machinery.
[10] J. Geiping, H. Bauermeister, H. Dröge, and M. Moeller. Inverting gradients - how easy is it to break privacy in federated learning? In *Advances in Neural Information Processing Systems*, 2020.
[11] J. Gerlach. Accelerated convergence in newton's method. *SIAM Rev.*, 36(2):272–276, 1994.
[12] R. Horn and C. Johnson. *Matrix Analysis*. Cambridge University Press, USA, 2nd edition, 2012.
[13] Z. Huang, R. Hu, Y. Guo, E. Chan-Tin, and Y. Gong. DP-ADMM: ADMM-based distributed learning with differential privacy. *IEEE Transactions on Information Forensics and Security*, 15:1002–1012, 2020.
[14] B. Jayaraman and D. Evans. Evaluating differentially private machine learning in practice. In *28th USENIX Security Symposium (USENIX Security 19)*, pages 1895–1912, Santa Clara, CA, 2019. USENIX Association.
[15] M. Kim, J. Lee, L. Ohno-Machado, and X. Jiang. Secure and differentially private logistic regression for horizontally distributed data. *IEEE Transactions on Information Forensics and Security*, 15:695–710, 2020.

[16] S. Li, C. Li, G. Chen, N. G. Bourbakis, and K.-T. Lo. A general quantitative cryptanalysis of permutation-only multimedia ciphers against plaintext attacks. *Signal Processing: Image Communication*, 23(3):212–223, 2008.

[17] K. Liu, C. Giannella, and H. Kargupta. An attacker's view of distance preserving maps for privacy preserving data mining. In *Proceedings of the 10th European Conference on Principles and Practice of Knowledge Discovery in Databases*, pages 297–308, Berlin, Germany, September 2006.

[18] K. Liu, H. Kargupta, and J. Ryan. Random projection-based multiplicative data perturbation for privacy preserving distributed data mining. *IEEE Transactions on Knowledge and Data Engineering*, 18(1):92–106, 2006.

[19] B. McMahan, E. Moore, D. Ramage, S. Hampson, and B. A. y. Arcas. Communication-Efficient Learning of Deep Networks from Decentralized Data. In *Proceedings of the 20th International Conference on Artificial Intelligence and Statistics*, volume 54 of *Proceedings of Machine Learning Research*, pages 1273–1282, 2017.

[20] H. B. McMahan, D. Ramage, K. Talwar, and L. Zhang. Learning differentially private recurrent language models. In *International Conference on Learning Representations*, 2018.

[21] L. Melis, C. Song, E. De Cristofaro, and V. Shmatikov. Exploiting unintended feature leakage in collaborative learning. In *2019 IEEE Symposium on Security and Privacy (SP)*, pages 691–706, 2019.

[22] P. Mohassel and Y. Zhang. SecureML: A system for scalable privacy-preserving machine learning. In *2017 IEEE Symposium on Security and Privacy (SP)*, pages 19–38, 2017.

[23] V. Nikolaenko, U. Weinsberg, S. Ioannidis, M. Joye, D. Boneh, and N. Taft. Privacy-preserving ridge regression on hundreds of millions of records. In *In 2013 IEEE Symposium on Security and Privacy*, pages 334–348, 2013.

[24] M. Pathak, S. Rane, and B. Raj. Multiparty differential privacy via aggregation of locally trained classifiers. In *Advances in Neural Information Processing Systems*, volume 23, 2010.

[25] M. Pilanci and M. J. Wainwright. Newton sketch: A near linear-time optimization algorithm with linear-quadratic convergence. *SIAM Journal on Optimization*, 27(1):205–245, 2017.

[26] M. J. Sheller, B. Edwards, G. Reina, and et al. Federated learning in medicine: facilitating multi-institutional collaborations without sharing patient data. *Sci Rep.*, 10(1):12598, 2020.

[27] M. Shen, X. Tang, L. Zhu, X. Du, and M. Guizani. Privacy-preserving support vector machine training over blockchain-based encrypted iot data in smart cities. *IEEE Internet of Things Journal*, 6(5):7702–7712, 2019.

[28] R. Shokri and V. Shmatikov. Privacy-preserving deep learning. In *2015 53rd Annual Allerton Conference on Communication, Control, and Computing (Allerton)*, pages 909–910, 2015.

[29] B. Strack, J. Deshazo, C. Gennings, J. L. Olmo Ortiz, S. Ventura, K. Cios, and J. Clore. Impact of hba1c measurement on hospital readmission rates: Analysis of 70,000 clinical database patient records. *BioMed research international*, 2014:781670, 04 2014.

[30] S. Truex, N. Baracaldo, A. Anwar, T. Steinke, H. Ludwig, R. Zhang, and Y. Zhou. A hybrid approach to privacy-preserving federated learning. In *Proceedings of the 12th ACM Workshop on Artificial Intelligence and Security*, AISec'19, page 1–11, New York, NY, USA, 2019. Association for Computing Machinery.

[31] Z. Wang, M. Song, Z. Zhang, Y. Song, Q. Wang, and H. Qi. Beyond inferring class representatives: User-level privacy leakage from federated learning. In *IEEE INFOCOM 2019 - IEEE Conference on Computer Communications*, pages 2512–2520, 2019.

[32] K. Wei, J. Li, M. Ding, C. Ma, H. H. Yang, F. Farokhi, S. Jin, T. Q. S. Quek, and H. V. Poor. Federated learning with differential privacy: Algorithms and performance analysis. *IEEE Transactions on Information Forensics and Security*, 15:3454–3469, 2020.

[33] Q. Yang, Y. Liu, T. Chen, and Y. Tong. Federated machine learning: Concept and applications. *ACM Trans. Intell. Syst. Technol.*, 10(2):1–19, 2019.

[34] S. Zhang, C. Tian, H. Zhang, J. Yu, and F. Li. Practical and secure outsourcing algorithms of matrix operations based on a novel matrix encryption method. *IEEE Access*, 7:53823–53838, 2019.

[35] Y. Zhang, X. Xiao, L. Yang, Y. Xiang, and S. Zhong. Secure and efficient outsourcing of PCA-based face recognition. *IEEE Transactions on Information Forensics and Security*, 15:1683–1695, 2020.

[36] Y. Zhang, R. Yu, M. Nekovee, Y. Liu, S. Xie, and S. Gjessing. Cognitive machine-to-machine communications: visions and potentials for the smart grid. *IEEE Network*, 26(3):6–13, 2012.

[37] L. Zhao and L. Chen. Sparse matrix masking-based non-interactive verifiable (outsourced) computation, revisited. *IEEE Transactions on Dependable and Secure Computing*, 17(6):1188–1206, 2020.

[38] W. Zheng, R. A. Popa, J. E. Gonzalez, and I. Stoica. Helen: Maliciously secure coopetitive learning for linear models. In *2019 IEEE Symposium on Security and Privacy (SP)*, pages 724–738, 2019.

[39] L. Zhu, Z. Liu, and S. Han. Deep leakage from gradients. In *Advances in Neural Information Processing Systems*, volume 32, 2019.